\documentclass[aps,onecolumn]{revtex4}  
\usepackage{epsfig,epsf}
\usepackage{graphicx}
\newcommand{\be}{\begin{equation}}
\newcommand{\ee}{\end{equation}}
\newcommand\beq{\begin{eqnarray}}
\newcommand\eeq{\end{eqnarray}}
\begin{document}

\title{Non-uniform chiral phase studied within the Polyakov NJL model}

\author{Tomasz L. Partyka}

\affiliation{Smoluchowski Institute of Physics, Jagellonian
University, Reymonta 4, 30-059 Krak\'ow, Poland}
 
\begin{abstract}
We consider how does the introduction of a Polyakov loop affects the spatially inhomogeneous quark condensate. The primary result of our work is that the existence of the spatially non-uniform chiral phase is confirmed within the Polyakov NJL model in a chiral limit. These findings are obtained both in a 3d-cutoff and in a Schwinger (proper time) regularization schemes. 
\end{abstract}

\maketitle

\section{Introduction}
\mdseries
The QCD phase diagram is one of the most interesting but still insufficiently explored topics in particle physics. Both theoretical and experimental efforts are currently taken to get a better understanding of the phase diagram of the strongly interacting matter. One of them are studies on the the spatially non-uniform chiral condensate.
Since the existence of the spatially non-uniform chiral phase in a dense quark matter was proposed \cite{dautry_bron_sad_jap}, many studies have been devoted to analyzing this subject \cite{NCh, nakano}.
People were interested in determining the range of baryon densities and temperatures for which the non-uniform chiral condensate is a ground state of a system. The Nambu - Jona-Lasinio model \cite{NJL}, as an effective theory, that describes chiral symmetry breaking was widely used to investigate the QCD phase diagram, and the results of the NJL model have been tested due to the change in regularization scheme and parameters choice. Only recently, the non-zero quark mass was included and the spataially non-uniform phase was studied not only in a chiral limit \cite{maedan,maedan2,partyka}. Nonetheless NJL model appeared to be very useful in analyzing a QCD phase diagram, it has a clear deficiency. NJL theory does not describes the significant aspect of QCD, what is a quark confinement. Hopefully, the Polyakov-NJL model was proposed as an extension of the original NJL theory \cite{PNJL, PNJL2}. In the Polyakov-NJL model the background gluon field is introduced and the expectation value of the Polyakov loop can be an indicator of a transition from a confined into a deconfined regions \cite{loop}. In the present paper, we study in the $\mu$ - T plane, how does the introduction of a Polyakov loop affects the spatially inhomogeneous condensate.

\section{Model} 
\mdseries
The initial lagrangian density of the PNJL model with two massless quarks has a form
\begin{eqnarray}
H=\int_x \overline{\psi}(i\gamma^\nu D_\nu +\mu\gamma_0-m)\psi+G\left[ (\overline{\psi}\psi )^2+(\overline{\psi}i\gamma_5\vec{\tau}\psi )^2\right] ,
\end{eqnarray}
where $\psi $ is the quark field and $\mu $ is the quark chemical potential. The color, flavor and spinor indices are suppressed. The vector
$\vec{\tau}$ is the isospin vector of Pauli matrices. The integration $\int_x=\int_0^\beta d\tau\int d^3x $,
where $\beta $ is the inverse temperature and the covariant derivative operator $D_\nu = \partial_\nu - i A_{\nu}(x)$, where $A_{\nu}(x)$ is a SU(3) gauge field.
The coupling constant $G$ describes interaction that is responsible for the creation of a 
quark-antiquark condensate. There is also an additional parameter $\Lambda$ which defines the energy scale below which
the effective theory applies. It is introduced through the regularization procedure.
We are working in the mean field approximation within the ansatz \cite{sad1}
\begin{eqnarray}
\langle\overline{\psi}\psi\rangle = -\frac{M}{2G}\cos\vec{q}\cdot\vec{x},\;\;\;
\langle\overline{\psi}i\gamma_5\tau^a\psi\rangle = -\frac{M}{2G}\delta_{a3}\sin\vec{q}\cdot\vec{x}
\end{eqnarray}
which describes three possible phases: the chiral uniform phase Ch ($\vec{q}=0, M\neq 0$), the non-uniform chiral phase NCh ($\vec{q}\neq 0, M\neq 0$) and the quark matter phase QM ($\vec{q}=0, M=0$). Because the PNJL model is described in detail in a vast literature on this subject \cite{ogol1, ogol2, poten},
 we only notice some basic assumptions. In our paper we use an effective potential of the form \cite{poten}
\begin{eqnarray}
{\cal{U}}(\Phi,\bar{\Phi},T)=T^{4}\biggl(-\frac{b_{2}(T)}{2}\Phi\bar{\Phi}-\frac{b_{3}}{6}(\Phi^{3}+\bar{\Phi}^{3})+\frac{b_{4}}{4}(\Phi\bar{\Phi})^{2}\biggr),\nonumber
\end{eqnarray}
\begin{eqnarray}
b_{2}(T)=6.75 -1.95\;\biggl(\frac{T_{0}}{T}\biggr) + 2.625\;\biggl(\frac{T_{0}}{T}\biggr)^{2} - 7.44\;\biggl(\frac{T_{0}}{T}\biggr)^{3},\;\;\;b_{3}=0.75,\;\;\;b_{4}=7.5,
\end{eqnarray}
where $\Phi$ and $\bar{\Phi}$ are the traced Polyakov loop and its conjugate. Following Ref. \cite{pawlow} we examine two scenarios for the parameter $T_{0}$. First, when $T_{0}$ is constatnt and equal 208 MeV. Second, when the additional dependence of a Polyakov critical temperature $T_{0}$ on a finite chemical potential is introduced,
\begin{eqnarray}
\label{mitem}
T_{0}=T_{\tau}e^{-1/(\alpha_{0}b(\mu))}, \;\;T_{\tau}=1.764 GeV, \;\;\alpha_{0}=0.304,\;\; b(\mu)=\frac{1}{6\pi}(11N_{c}-2N_{f})-\frac{16}{\pi}N_{f}\frac{\mu^{2}}{T_{\tau}^{2}}.
\end{eqnarray}
Next, we assume that the spatial term of a gluon gauge field is vanishing and that the temporal component $A_{0}$ is constant. With above assumptions, the traced Polyakov loop $\Phi$ is expressed by the gluon filed $A_{0}$ in the form
\begin{eqnarray}
\Phi=Tr L/N_{c},\;\; L=\exp(-\beta A_{0}).
\end{eqnarray}
Because we stay in a mean field limit, the expectation values $<\Phi>$ and $<\bar{\Phi}>$ are equal to each other and real \cite{fifi}. To determine the ground state of the system we calculate the thermodynamic potential
\be
\Omega_{0}=\frac{M^{2}}{4G}-\frac{T}{V}\ln \int D\overline{\psi}D\psi\ \exp\biggl(\int_{0}^{\beta}\int{d^{3}x}{\overline{\psi}S_{0}^{-1}\psi}\biggr)+{\cal{U}}(\Phi,\bar{\Phi},T),
\ee
where
\begin{eqnarray}
S^{-1}_{0}=\left[i\gamma^{\nu}(D_{\nu}-\frac{1}{2}i\gamma_{5}\tau_{3}q^{\nu})+\mu\gamma_{0}-M\right].
\end{eqnarray}

Using the standard method one can obtain an explicite form of the potential $\Omega_{0}$
\begin{eqnarray}
\Omega_{0}&=& \frac{\;M^2}{4G} 
\;-\;6\sum_{s=\pm}\int\frac{d^3k}{(2\pi)^3} (E_s)\;+\;{\cal{U}}(\Phi,\bar{\Phi},T)\nonumber \\ 
&-&2T\sum_{s=\pm}\int\frac{d^{3}k}{(2\pi)^{3}}\left[\ln\left(1+3\;(\bar{\Phi}+\Phi\exp\biggl(-\frac{(E_{s}+\mu)}{T}\biggr))\exp\biggl(-\frac{(E_{s}+\mu)}{T}\biggr)+ \exp\biggl(-3\;\frac{(E_{s}+\mu)}{T}\biggr)\right)\right]\nonumber \\
&-&2T\sum_{s=\pm}\int\frac{d^{3}k}{(2\pi)^{3}}\left[\ln\left(1+3\;(\Phi+\bar{\Phi}\exp\biggl(-\frac{(E_{s}-\mu)}{T}\biggr))\exp\biggl(-\frac{(E_{s}-\mu)}{T}\biggr)+\exp\biggl(-3\;\frac{(E_{s}-\mu)}{T}\biggr)\right)\right]
\end{eqnarray}
where  
\be
E_\pm=\sqrt{\vec{k}^2+M^2+\frac{\vec{q}^{\,2}}{4}\pm \sqrt{(\vec{q}\cdot\vec{k})^2+M^2\vec{q}^{\,2}}}.
\ee
The PNJL model is a nonrenormalizable theory, that is why, we apply a regularization procedure that finally defines the model. We use a 3d cutoff scheme and a Schwinger (proper time) scheme. Parameters of the model are uniquely fixed to reproduce the values of the physical quantities in the vacuum, the pion decay constant $f_{\pi}=$ 93 MeV and the quark condensate density $<\overline{u}u>=<\overline{d}d>= -(250 MeV)^{3}$ \cite{klev, tlms}. After these procedure, thermodynamic potential $\Omega_{0}$ is a fuction of a chiral order parameter M, an absolute value of the wave vector $\vec{q}$, a traced Polyakov loop $\Phi$, a quark chemical potential $\mu$ and temperature T. For constant $\mu$ and T we numerically solve the mean field equations
\begin{eqnarray}
\frac{\partial \Omega_{0}}{\partial M}=\frac{\partial \Omega_{0}}{\partial q}=\frac{\partial \Omega_{0}}{\partial \Phi}=0.
\end{eqnarray}

\section{Results} 
\subsection{Phase Diagram}
Our aim is to resolve whether in the framework of the PLNJ model, the spatially non-uniform condensate could exists and indicate in the $\mu$ - T plane the region where possibly the NCh phase is a global minimum. 
We also show, how the particular parameters of the thermodynamic potential $\Omega_{0}$ depend on $\mu$ and T. 
We begin this analysis with a brief introduction to the appearance of the phase diagram of the NJL model (with two massless quarks) \cite{sad}. In Fig. 1 we can see the illustration of a $\mu$ - T phase diagram with pre-fixed $q=0$ (left panel) and with $q \neq0$ (right panel). 
Note that introducing a possibility of an existence of the spatially inhomogeneous phase results in a distinct change of a critical chemical potential (transition between chiral and QM phase). The point of a chiral phase transition (at zero temperature) is shifted about 125 MeV towards the larger baryon density. The line of a transition from the Ch into the NCh region (right panel) is slightly moved to the left in comparison with a transition line from the Ch into the QM phase (left panel). The phase transitions between Ch/NCh, NCh/QM are first order but the transition line between Ch/QM (right panel of Fig. 1) is a smooth crossover. 
The phase transition between Ch/QM (left panel of Fig. 1) is  of the first order untill the critical point (marked by the black dot).

\begin{figure}[h]
\centerline{\epsfxsize=6.95 cm \epsfbox{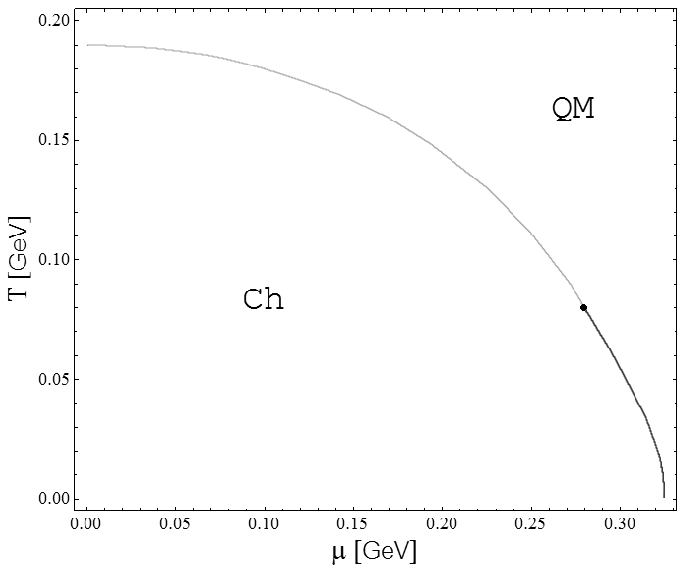} \epsfxsize=9.03 cm \epsfbox{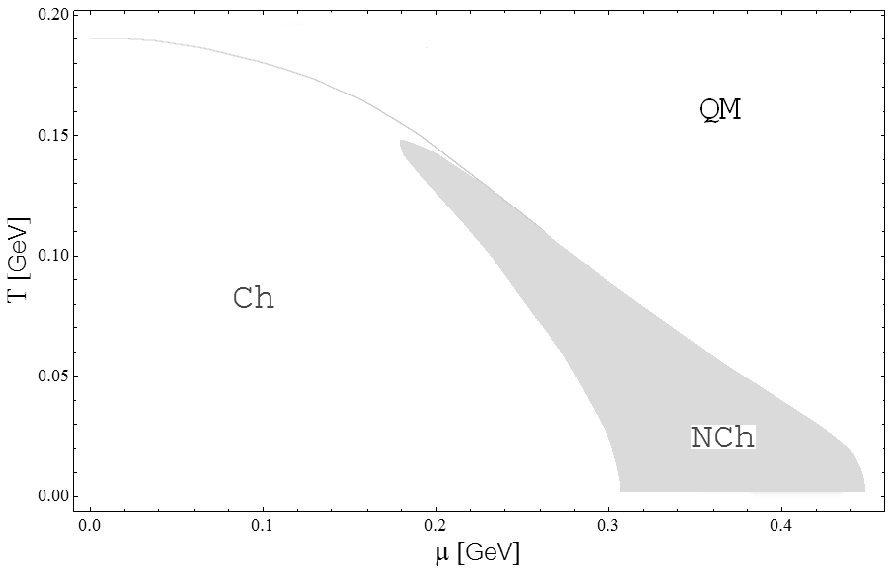}}
\caption{The phase diagrams of a NJL model (with two massless quarks) in a 3d cut-off regularization scheme, G$\Lambda^2$=2.14, $\Lambda=$653 MeV. The critical end point (black dot) on a left panel is located at $\mu=$0.28 GeV, T =0.08 GeV.}
\end{figure} 

\begin{figure}[h]
\centerline{\epsfxsize=6.95 cm \epsfbox{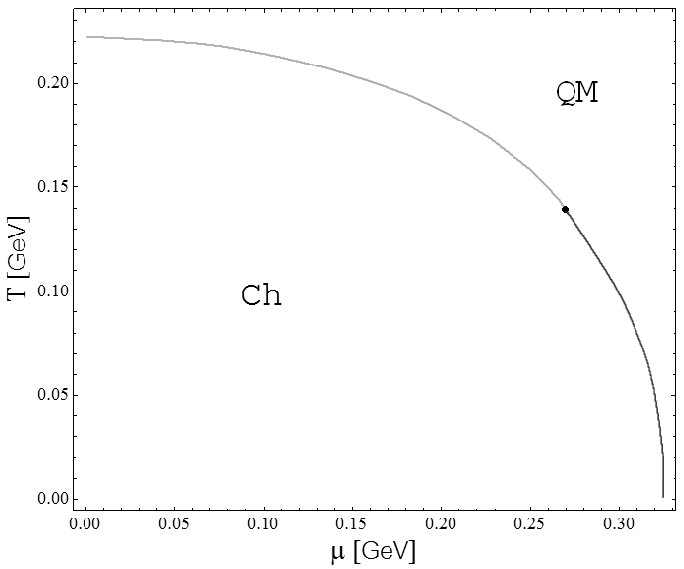} \epsfxsize=9.03 cm \epsfbox{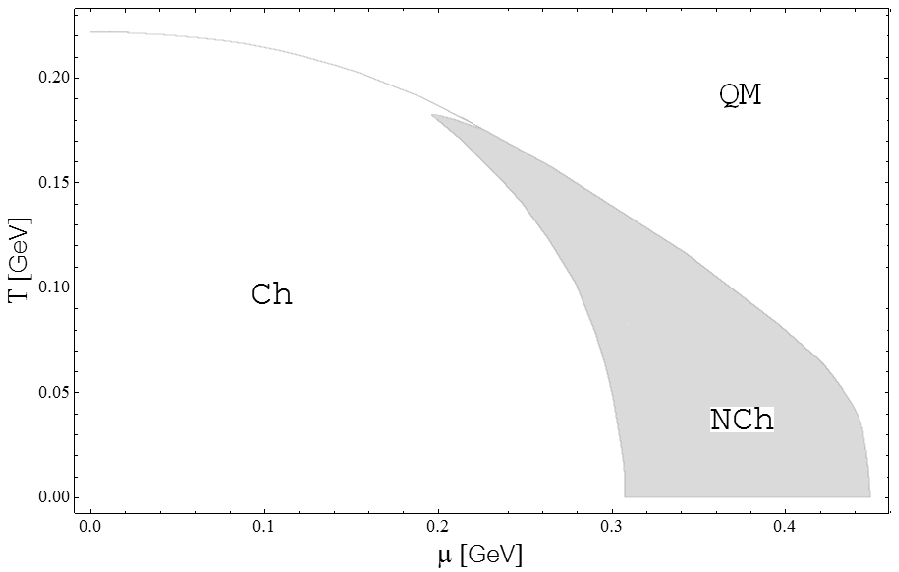}}
\caption{The phase diagrams of a PNJL model (with two massless quarks) in a 3d cut-off regularization scheme, G$\Lambda^2$=2.14, $\Lambda=$653 MeV, $T_{0}=$208 MeV. The critical end point (black dot) on a left panel is located at $\mu=$0.27 GeV, T =0.14 GeV.} 
\end{figure}
Now, we consider the same diagrams but obtained in the Polyakov NJL model. In Fig. 2, there are presented phase diagrams, without NCh condensate (left panel) and with NCh phase (right panel). As one can expected, because the Polyakov potential is proportional to the fourth power of temperature, the introduction of an effective potential for $\Phi$, affects diagrams mainly in a large temperature region, while at T = 0, the critical chemical potentials remain unchanged. As it turns, there is a region where the ground state of the system is realized by the spatially non-uniform condensate (gray area). The nature of phase transitions in Fig. 2 is the same as previously. On the left panel of Fig. 2, we see that temperature of a chiral crossover at $\mu=$ 0 is now $T_{c}=$ 221 MeV. It is a consequence of an interplay between the Polyakov effective potential  ${\cal{U}}(\Phi,\bar{\Phi},T)$ and the rest part of the thermodynamic potential $\Omega_{0}$ proportional to temperature. The chiral crossover and the deconfinement transition are distinct phenomena, however they are linked with each other. In the presence of dynamical quarks, the expectation value of a traced Polyakov loop plays a role of an indicator of a deconfinement (not an exact order parameter). We can indicate a region of a very rapid change in $\Phi$ and linked that region with a deconfinement transition.
In a scenario when the parameter $T_{0}$ of a Polyakov potential ${\cal{U}}$ is constant and equal 208 MeV, the coincidence between a chiral and a deconfinement transitions is valid only for small chemical potentials and large temperatures. Apart from these region, the deconfinement transition line goes after the  chiral one, what is an unphyssical efect \cite{pawlow} and that is why we do not present the deconfinement line in Fig. 2. The partial solution to that situation is suggested in Ref \cite{pawlow}. Additional $\mu$-dependence of a critical temperature $T_{0}$ is introduced (\ref{mitem}). The phase diagram obtained in this latter scenario is presented in Fig. 3. $T_{0}(\mu)$ is getting smaller with growing chemical potentaial, what results in a fact that a deconfined transition (marked by the dashed line) precedes a chiral transition and this relation is valid untill the region of  very large $\mu$ and small temperature.

\begin{figure}[h]
\centerline{\epsfxsize=8.22 cm \epsfbox{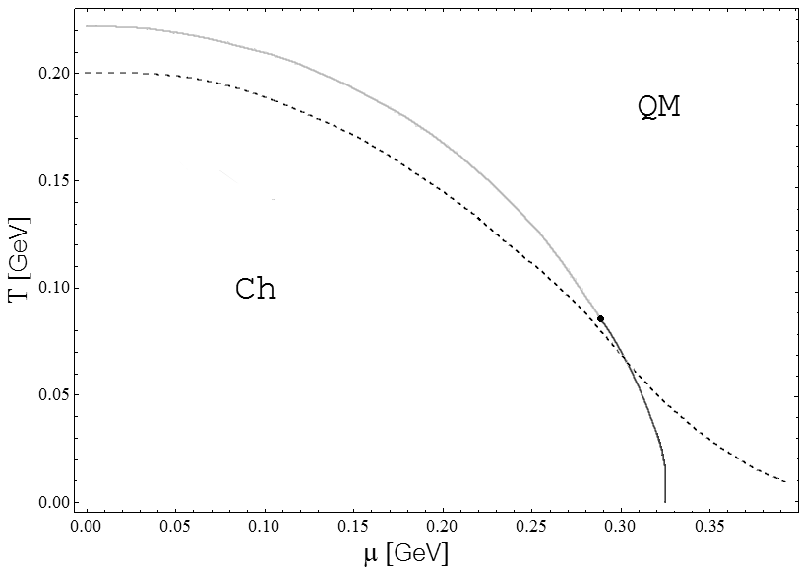} \epsfxsize=9.03 cm \epsfbox{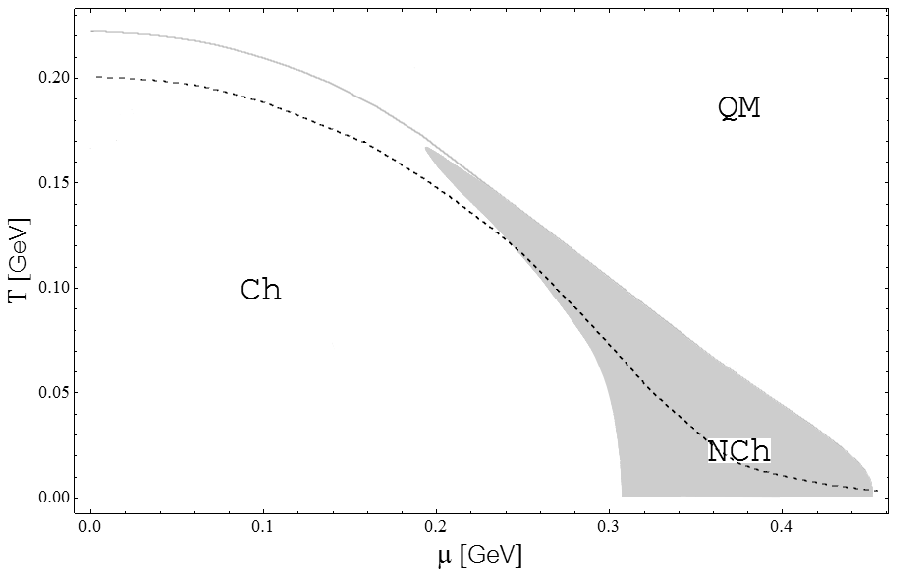}} 
\caption{The phase diagrams of a PNJL model (with two massless quarks) in a 3d cut-off regularization scheme, G$\Lambda^2$=2.14, $\Lambda=$653 MeV, $T_{0}=T_{0}(\mu)$. The critical end point (black dot) on a left panel is located at $\mu=$0.288 GeV, T =0.085 GeV. Dotted line denotes the deconfinement transition.}
\end{figure} 
The outcomes of the PNJL model are strongly dependent on the choice of parameters and regularization procedure. Therefore, we also examine the results obtained in a Schwinger regularization.   In Fig. 4, the phase diagram obtained in a proper time scheme is presented. $T_{0}=$208 MeV on the left panel of Fig. 4 and on the right, $T_{0}$ is a function of $\mu$. The values of a critical chemical potential and temperature are different than in a 3d cutoff scheme. The chiral crossover and the deconfinement transition at zero baryon density almost coincide, while in a 3d cutoff scheme, the distance between them was about 20 MeV. Despite these quantitative differences, the main result is common and the region of an existence of the NCh condensate is marked in gray. 

\begin{figure}[h]
\centerline{ \epsfxsize=7.27 cm \epsfbox{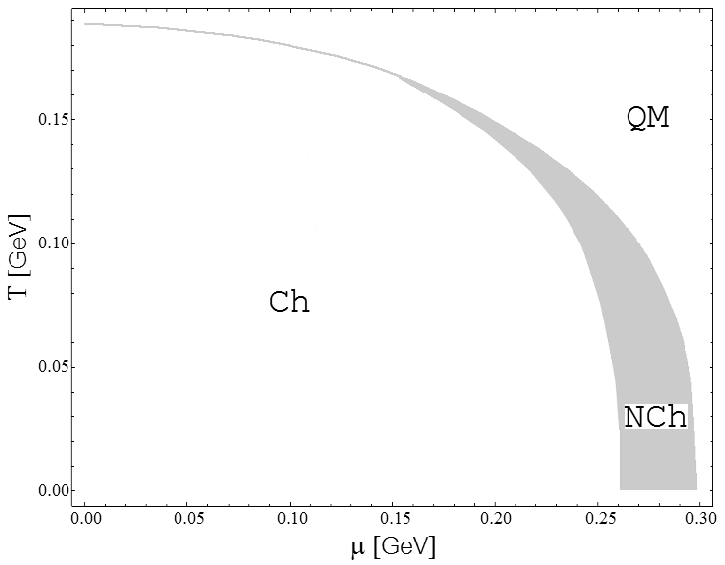} \epsfxsize=8.31 cm \epsfbox{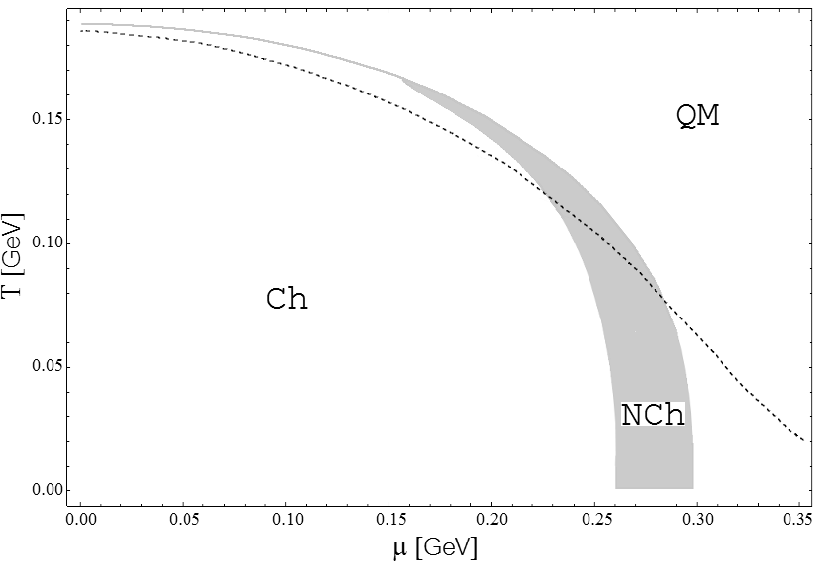}}
\caption{The phase diagrams of a PNJL model (with two massless quarks) in a Schwinger regularization scheme, G$\Lambda^2$=3.78, $\Lambda=$1086 MeV. On the left panel $T_{0}=$208 MeV, on the right panel $T_{0}=T_{0}(\mu)$. Dotted line denotes the deconfinement transition.}
\end{figure}
\begin{figure}[h]
\vspace{1 cm}
\centerline{\epsfxsize=9.9 cm \epsfbox{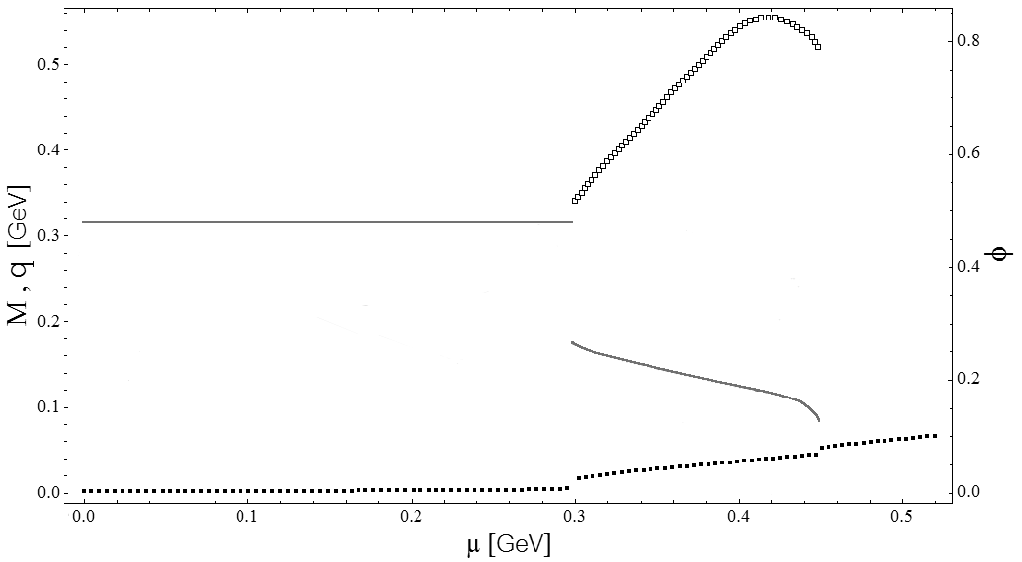} \epsfxsize=9.04 cm \epsfbox{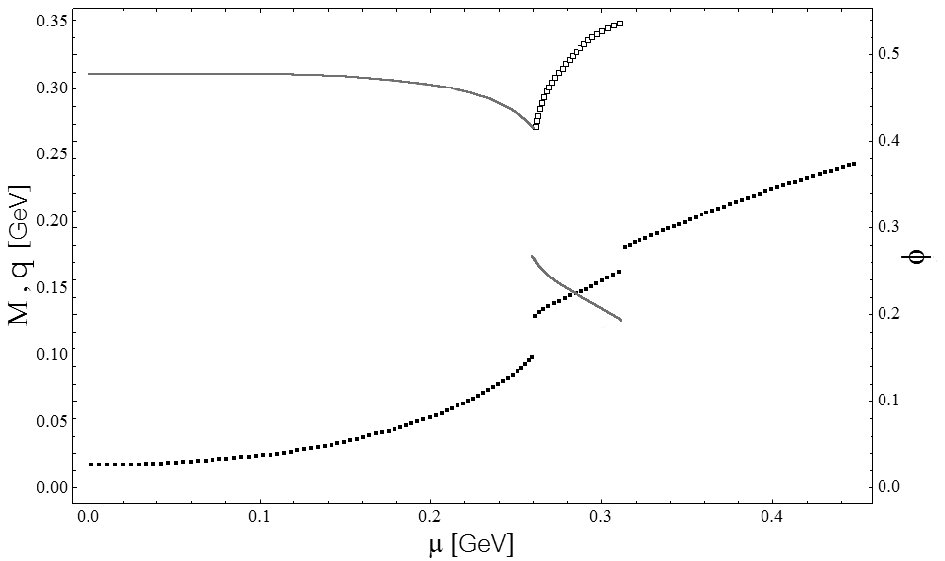}}
\caption{The dependence of M (gray line), q (open boxes) and $\Phi$ (dotted line) on $\mu$ at one plot in a 3d cut-off regularization scheme, G$\Lambda^2$=2.14, $\Lambda=$653 MeV, $T_{0}=$208 MeV. Left panel T = 20 MeV, right panel T = 120 MeV.}
\end{figure}
\subsection{Parameter Dependences}
 \begin{figure}[h]
\centerline{\epsfxsize=9.04 cm \epsfbox{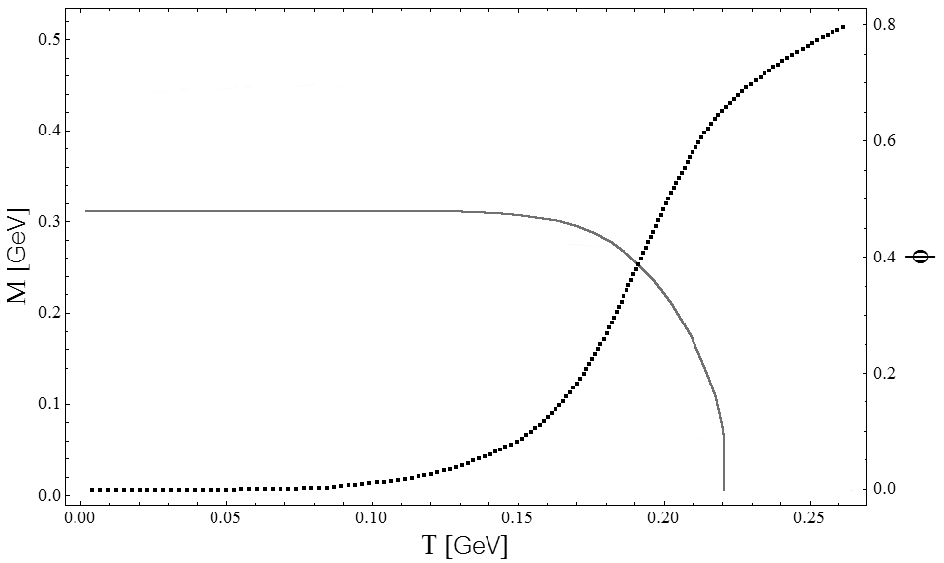} \epsfxsize=9.04 cm \epsfbox{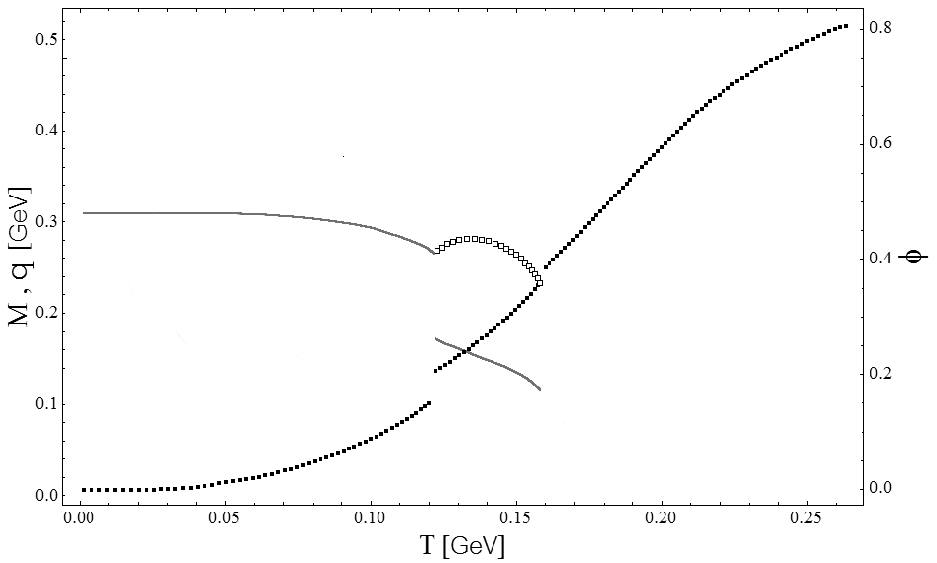}}
\caption{The dependence of M (gray line), q (open boxes) and $\Phi$ (dotted line) on T at one plot in a 3d cut-off regularization scheme, G$\Lambda^2$=2.14, $\Lambda=$653 MeV, $T_{0}=$208 MeV. Left panel $\mu$ = 0 MeV, right panel $\mu$ = 260 MeV.}
\end{figure}
 \begin{figure}[h]
\centerline{\epsfxsize=9.04 cm \epsfbox{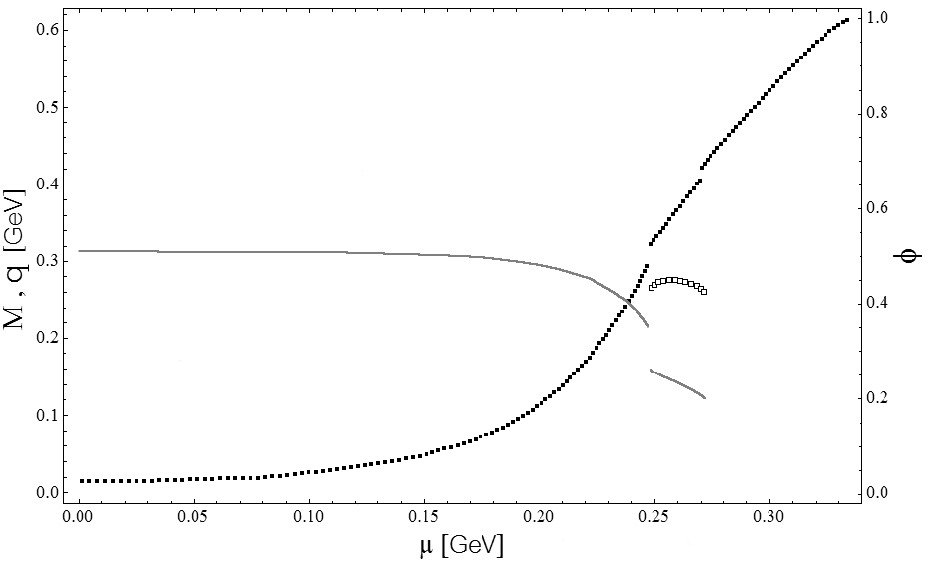} \epsfxsize=9.04 cm \epsfbox{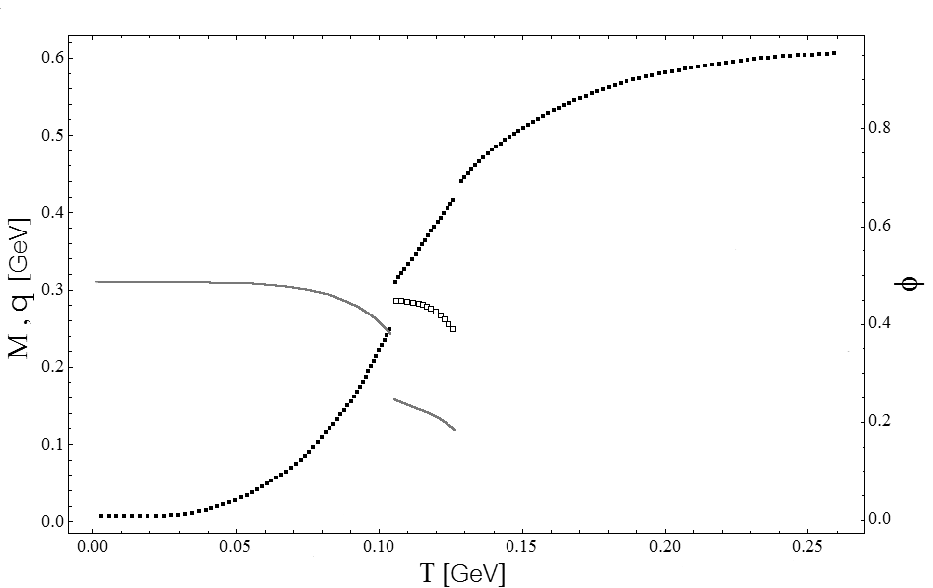}}
\caption{The dependence of M (gray line), q (open boxes) and $\Phi$ (dotted line) at one plot in a 3d cut-off regularization scheme, G$\Lambda^2$=2.14, $\Lambda=$653 MeV, $T_{0}=T_{0}(\mu)$. Left panel T = 120 MeV, right panel $\mu$ = 260 MeV.}
\end{figure}
A nice way to show the behavior of the various parameters of the thermodynamic potential $\Omega_{0}$ is to perform several cross-sections through the full phase diagram. In Fig. 5, we present the dependence of a quark constituent mass M, an absolute value of a wave vector q and a traced Polyakov loop $\Phi$ on $\mu$, for respectively T = 20 MeV (left panel) and T = 120 MeV (right panel). 
We see, that this dependence is very similar to that obtained in the original NJL model \cite{sad, tlms}. Both a transition from the uniform chiral to the non-uniform chiral and a transition from the non-uniform phase to the quark matter phase are of the first order. The value of $\Phi$ is gradually growing with an increase of a chemical potential. It is difficult to find any physical interpretation for a discontinuity of the $\Phi$ value in a chemical potentials corresponding to Ch/NCh and NCh/QM transitions. Nevertheless, we remember that in our case, the traced Polyakov loop $\Phi$ is only an indicator of a deconfinement, and so we are interested mostly in the qualitative results.
In Fig. 6, the additional analysis is presented. The dependence of M, q and $\Phi$ on temperature is examined for a constant chemical potential. On the left panel of Fig. 6, one can observe a mentioned earlier merging of a chiral crossover and a deconfinement transition. Region of a very rapid change in $\Phi$ coincides with the disappearance of M. On the right  panel of Fig. 6 ($\mu=$ 260 MeV) the growth of $\Phi$ is more gradual. Again, there are discontinuities of a $\Phi$ value, one associated to the Ch/NCh transition and the other to the NCh/QM transition. On the left panels of Fig. 5, 6, (scenario when $T_{0}=$ 208 MeV) it is visible that while the coincidence between a chiral and a deconfinement transitions makes sense for small chemical potentials and high temperatures, in the rest part of a diagram, this agreement is not entitled. Introduction of a $\mu$-dependence of a Polyakov critical temperature causes that the region of a rapid growth in $\Phi$ appears for smaller chemical potentials and the growth is much steeper. It is visible in Fig. 7, where the M, q and $\Phi$ values are plotted for respectively T = 120 MeV left panel and $\mu=$260 MeV right panel.
\newpage
\section{Conclusions} 
The primary result of our work is that the existence of the spatially non-uniform chiral phase is confirmed within the Polyakov NJL model in a chiral limit. These findings are obtained both in a 3d cutoff regularization and in a Schwinger (proper time) schemes. We have examined two scenarios of the critical Polyakov temperature $T_{0}$ dependence. One, when $T_{0}$ is constant and equal 208 MeV, the other, when the $T_{0}$ value is lowering with growing $\mu$. For both of these scenarios, in comparison to the original NJL model, the temperature of a chiral crossover in $\mu=$ 0 changes from 190 to 221 MeV (3d cutoff) and from 152 to 189 MeV (proper time), while the critical chemical potentials at T = 0 between Ch/NCh and NCh/QM remain unchanged. As a consequence the line of a transition from chiral phases into the quark matter phase is always above the same line obtained in the original NJL model. This effect is stronger for constant $T_{0}$. Introduction of a $T_{0}(\mu)$ dependence causes a better agreement of a chiral and a deconfinement transitions. 
The future work that should be done to get a better understanding of the PNJL phase diagram is to concider simultaneously the non-zero current quark mass and inhomogeneous condensate. In the present paper we stay in a mean field approximation, the analysis of a diagram away from this limit is also worth consideration.\\
\\
\textbf{Acknowledgement:} 
I would like to thank Professor Mariusz Sadzikowski for many interesting discussions and comments.

\end{document}